# Image classification using collective optical modes of an array of nanolasers

GIULIO TIRABASSI,[1,+] JI KAIWEN,[2,x] CRISTINA MASOLLER,[1,*] AND ALEJANDRO M. YACOMOTTI [2,v]

[1]*Departament de Fisica, Universitat Politecnica de Catalunya, Rambla Sant Nebridi 22, 08222 Terrassa, Barcelona, Spain*
[2]*Centre de Nanosciences et de Nanotechnologies, CNRS, Universite Paris-Sud, Université Paris-Saclay, 10 Boulevard Thomas Gobert, 91120 Palaiseau, France*
[+]*giulio.tirabassi@upc.edu*
[x]*ji.kaiwen@c2n.upsaclay.fr*
[*]*cristina.masoller@upc.edu*
[v]*alejandro.giacomotti@c2n.upsaclay.fr*

**Abstract:** Recent advancements in nanolaser design and manufacturing open up unprecedented perspectives in terms of high integration densities and ultra-low power consumption, making these devices ideal for high-performance optical computing systems. In this work we exploit the symmetry properties of the collective modes of a nanolaser array for binary image classification. The implementation is based on a 8x8 array, and relies on the activation of a collective optical mode of the array—the so-called "zero mode"—, under spatially modulated pump patterns. We demonstrate that a simple training strategy allows us to achieve an overall success rate of 98% in binary image recognition.

## 1. Introduction

Photonic artificial neural networks (ANNs) are sparking a revolution in artificial intelligence (AI) systems because they have the potential of being much faster and energy efficient than current silicon technology [1–3]. In these Big Data days where datacenters consume enormous amounts of power and the increase of computing performance, based on the increasing number of transistors, is reaching fundamental miniaturization limits (Moore's law), faster and more energy efficient AI systems are urgently needed.

Impressive advances have been made in improving the performance of photonic ANNs, designing hardware that mimics neural synapses, developing efficient training methods, expanding the number of nodes and integrating them into silicon chips [4–10]. As a recent example, a specialized photonic processor has been demonstrated with a performance that is 2-3 orders of magnitude higher than the equivalent digital electronic processor [11]. Moreover, a survey of the best-in-class integrated photonic devices has reported that silicon photonics can compete with the best-performing electronic ANNs, reaching sub-pJ per MAC (multiply-accumulate operations) and foresees, for sub-wavelength photonics, performances of few fJ/MAC [12].

ANNs require activation functions which, in photonics, typically rely on strongly nonlinear mechanisms such as optical bistability or saturable absorption [13, 14]. Usual implementations with coherent nanophotonic circuits require additional nonlinear materials, such as graphene layers, that are integrated in a second stage [13], which increase the technological complexity. Alternatively, a laser cavity features a natural activation function in the form of gain —instead of absorption— saturation at the laser threshold. Moreover, in semiconductor quantum wells or quantum dots, cross-gain saturation leads to strong nonlinear mode competition, providing a mode selectivity mechanism that we attempt to exploit in this work. In a different approach, laser-based ANNs have also been used for simulating spin Hamiltonians [15, 16], multimode

VCSELs have been shown to enable parallel ANNs [17], and a laser with intracavity spatial light modulator (SLM) has been implemented as a rapid solver for the phase retrieval problem [18].

A most appealing candidate for laser-based integrated ANNs is a nanolaser. Nanolasers have huge potential for becoming the key building blocks of scalable, photonic computing systems, able to provide both high-performance and ultra-low energy consumption [19]. Remarkably, nanolaser technology has experimented a huge progress in the recent years, enabling dense integration of laser nanosources on photonic microchips. Today, a myriad of cavity designs and materials come to maturity for realizing nanolaser arrays, with advantages in terms of ultracompact footprints, low thresholds, and room-temperature operation [20]. Among these technological platforms we can identify photonic crystal, metallo-dielectric and coaxial-metal nanolasers, together with plasmonic lasers or spasers; in addition, quantum-dot-based micropillar lasers feature microcavity properties that make them promising for applications in, e.g., reservoir computing [21]. The common key physical mechanism of both micro and nanolasers leading to low energy consumption is their high quality factors (Q) combined with potentially high spontaneous emission factors ($\beta$), that ultimately bring the device operation deep into the thresholdless and few-photon regime. Furthermore, lasers with micro/nanocavities have two additional assets: i) evanescent coupling between neighboring cavities may lead to strong optical coupling, and consequently a micro/nano laser array can be robust against fabrication imperfections; and ii) intercavity coupling can be tailored by design, with unprecedented control in photonic crystal platforms, that enable choosing both the magnitude and sign of the coupling parameters [22–24].

In this work we propose a photonic ANN based on an array of nanolasers, which is able to process and classify two classes of images. As a proof-of-concept demonstration, we process a database containing images of handwritten numbers. The input image (input layer) is encoded into a spatially modulated pump beam, which excites the array and enables full connectivity between the inputs and the nanolasers (hidden layers). The result of the classification task —to fix the ideas, *yes* if the image has a handwritten zero, *no* otherwise— is read from the optical spectrum of the light emitted by the nanolaser array, which constitutes the output decision layer (the emitted modes can be considered the "nodes" of the output layer). Consequently, we can select a set of modes such that they are activated —i.e., pumped above laser threshold— only for a given set of input images (handwritten zeros), and they remain off —i.e., below threshold— otherwise. The key idea is that the spatial profile of the pump is appropriately designed such that, when it encodes a specific type of image, it activates a particular set of modes of the array. Therefore, from the analysis of the frequencies of the emitted modes we can infer the class of the image (i.e., a zero or a non-zero digit) that is encoded in the spatial pattern of the pump.

Similarly to the optical computing framework demonstrated in [25], based on nonlinear mode interactions in a multimode fiber, in our implementation the activation function is non-local as self and cross-gain saturation result in thresholding functions with high mode competition and selectivity. A crucial advantage of our setup is the use of nanolasers, which have ultra-low power consumption and very small footprint.

This work is organized as follows. Section 2 presents the model used to describe the nanolaser array, and introduces the sublattice symmetry that allows to describe the evanescently coupled nanolaser array as an optical bi-layer. In Sec. 2 we also introduce the definition of a particular type of collective mode, known as zero-mode. Section 3 describes the use of zero-mode lasing for binary classification, i.e., to determine whether an image that is encoded in the spatial profile of the pump corresponds (yes/no) to a particular digit. The dataset used, the machine learning algorithm developed to optimize the system's performance, and its implementation are described in Secs. 4 and 5. Sections 6, 7 and 8 present the results obtained, the discussion and the conclusions.

## 2. Model

We consider a two-dimensional nanocavity array (shown in blue in Fig .1), that consists of $m \times n$ cavities in the $x$ and $y$ directions, respectively. For simplicity, we only consider nearest-neighbor coupling and further assume that the resonance frequencies are the same for all the cavities (without loss of generality, we set them to be zero). The system is described by the following coupled-mode equations:

$$i\frac{da_{m,n}}{dt} = \kappa_x(a_{m-1,n} + a_{m+1,n}) + \kappa_y(a_{m,n-1} + a_{m,n+1}) + i(g_{m,n} - \gamma)a_{m,n}. \qquad (1)$$

Here, $a_{m,n}$ are the field amplitudes, $g_{m,n}$ stand for the gain rates, which are the elements of the matrix that represents the spatial pump pattern ($\boldsymbol{P} = \{g_{m,n}\}$), and $\gamma$ is the optical loss rate, which is assumed the same for all the cavities; $\kappa_x$ and $\kappa_y$ are the coupling rates in $x$ and $y$ directions, respectively.

For the numerical calculations, we normalize the time in Eq. (1) to the timescale $T = 0.2\lambda^2/\pi c\Delta\lambda$, where $\lambda$ is the cavity-resonance frequency, $c$ is the speed of light, and $\Delta\lambda$ is the resonance linewidth. For typical photonic crystal nanolasers with $\lambda = 1550$ nm and quality factors of $Q \sim 4000$, $\Delta\lambda \approx 0.4$ nm, and hence $T \approx 1.27$ ps. We then re-scale all the rates as $\gamma \rightarrow \gamma T$, $g_{m,n} \rightarrow g_{m,n}T$, $\kappa_x \rightarrow \kappa_x T$, and $\kappa_y \rightarrow \kappa_y T$. Dissipation and coupling parameters can be modified through Q-factor and evanescent coupling engineering, respectively, which provides important degrees of freedom in the coupled cavity design. Throughout this work we will fix $\gamma = 0.2$ and $\kappa_x = \kappa_y = 1$, consistent with standard coupled photonic crystal nanolaser geometries (see Sec. 5).

Next, we rewrite these equations in Dirac notation

$$i\frac{\partial |\psi\rangle}{\partial t} = \boldsymbol{H}|\psi\rangle, \qquad (2)$$

where

$$\boldsymbol{H} = \sum_{m,n} \left((\kappa_x |a_{m,n}\rangle\langle a_{m+1,n}| + \kappa_y |a_{m,n}\rangle\langle a_{m,n+1}| + h.c.) + i(g_{m,n} - \gamma)|a_{m,n}\rangle\langle a_{m,n}|\right). \qquad (3)$$

Here $h.c.$ stands for the Hermitian conjugate and

$$|\psi\rangle = (a_{1,1}, a_{1,2}, \cdots, a_{m,n})^T, \qquad (4)$$

is the wavefunction, that is a vector containing the $m \times n$ complex amplitudes.

A 2D cavity array as modelled by Eq. (1) may feature important symmetries such as the chiral —also know as sublattice— symmetry, where $\boldsymbol{H}$ (not necessarily Hermitian) anti-commutes with a unitary operator. In the case of non-Hermiticity provided by a gain/loss distribution in a coupled cavity Hamiltonian verifying $(\kappa_x, \kappa_x) \in \mathbb{R}$, the prevailing underlying symmetry is the non-Hermitian particle-hole (NHPH) symmetry [26, 27]. In this case, the Hamiltonian in Eq. (3) satisfies the anticommutation relation $\boldsymbol{HC\mathcal{T}} = -\boldsymbol{C\mathcal{T}H}$, where $\boldsymbol{C}$ is the unitary operator and $\mathcal{T}$ is the time reversal operator [26].

Both chiral and particle-hole symmetries apply to arrays that can be decomposed in two sublattices $A$ and $B$, where coupling only takes place between cavities from different sublattices. In the case of the 2D square lattice displayed in Fig. 1, the sublattices $A$ and $B$ are defined by the schematized checkerboard (red and blue sites). Importantly, in the context of ANNs, because of the sublattice decomposition, a 2D cavity array with nearest neighbour coupling can be mapped to a bi-layer or bipartite network (also shown in Fig. 1). Consequently, with nearest neighbour coupling, the 2D nanolaser array is a bi-layer ANN, in which both layers are coupled to an input matrix $\boldsymbol{I}$, and to an output layer that is the eigenmode spectrum, $\mathfrak{R}(\epsilon)$ (Fig. 1, right column).

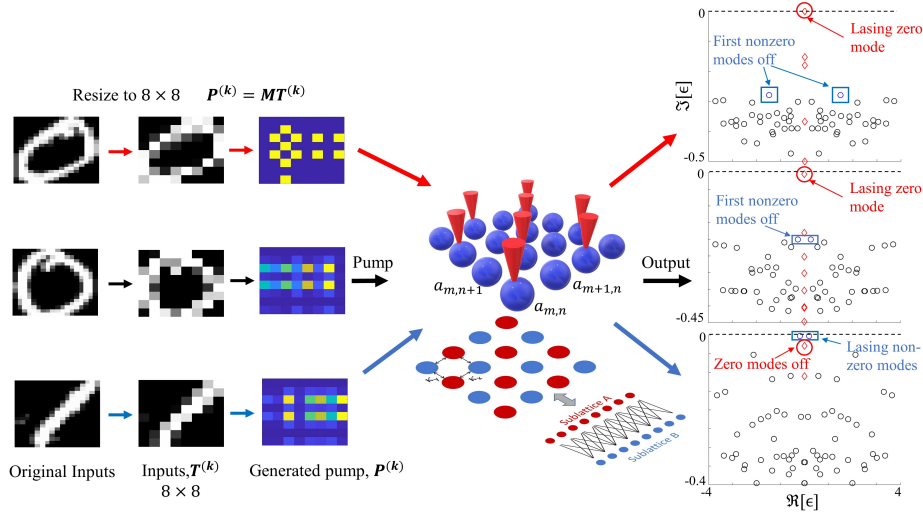

Fig. 1. Examples of handwriting digital identification. The first column shows the original input images, the second column, the resized images and third column, the pump patterns generated by a linear transformation. The eigenvalues are shown in the right column, where the dash lines indicate the lasing threshold. We see that the zero mode is above threshold for the two zero digit images (red and black arrows) while it is below threshold for the non-zero digit image (blue arrow). The nanocavity array (blue spheres) can be decomposed in two subalattices (A in red, B in blue) that constitute the hidden layers of the photonic ANN.

To investigate the properties of the NHPH symmetry, we diagonalize the Hamiltonian, $H|\phi_i\rangle = \epsilon_i|\phi_i\rangle$, where $\epsilon_i$ and $|\phi_i\rangle$ are the eigenvalues and the eigenvectors respectively. According to the anticommutation relation, we have that

$$H(CT|\phi_i\rangle) = -CT(H|\phi_i\rangle) = -CT\epsilon_i|\phi_i\rangle = -\epsilon_i^*(CT|\phi_i\rangle). \quad (5)$$

Therefore, $CT|\phi_i\rangle$ is also an eigenvector of $H$ with eigenvalue $-\epsilon_i^*$. We can denote the new eigenvector and eigenvalue as $|\phi_j\rangle = CT|\phi_i\rangle$ and $\epsilon_j = -\epsilon_i^*$, respectively. Consequently, the NHPH symmetry possess two phases, the symmetry-broken phase with $\epsilon_j = -\epsilon_i^*, i \neq j$, and the symmetric phase with $\epsilon_i = -\epsilon_i^*$, which implies that the NHPH-symmetric modes are *zero-modes*, defined by the condition $\Re[\epsilon_j] = 0$, and they are simultaneous eigenvectors of $H$ and $CT$. Importantly, two eigenvalues that initially have different energies, may, upon variation of the pump parameter, collide on the imaginary axis at an exceptional point and create a pair of NH zero-modes: this phenomenon is called the *spontaneous restoration* of the NHPH symmetry.

Zero-modes have captivated the attention of the scientific community because of the revolutionary concept of Majorana bound states. They constitute their own anti-particles and host non-Abelian braiding statistical properties, a most promising approach for fault tolerant topological quantum computation. In photonics, non-Hermitian zero modes have been recently demonstrated in an array composed by three nanocavities [27].

### 3. Binary classification using the zero-modes

The symmetric phase can be observed when the array is pumped in a selective manner [26].

Unlike the zero-modes protected by chiral symmetry that satisfy $\epsilon_i = -\epsilon_i$, and thus, are restricted to $\epsilon_i = 0$, the zero-modes warranted by NHPH symmetry satisfy $\Re[\epsilon_i] = 0$, therefore they are free to move on the imaginary axis. Consequently, the NH zero-modes are robust, and one can manipulate them by controlling the pump pattern. Upon pumping the array with an appropriate spatial pattern, a zero-mode can reach the lasing threshold, $\Im[\epsilon_i] = 0$, before the other modes.

On the other hand, if the spatially distributed pump cannot spontaneously restore the symmetric phase, a non-zero mode will lase. Thus, the pump patterns can be classified into two groups according to their ability to excite a zero-mode: the ones that can can lead to zero-mode lasing and the ones that cannot. Therefore, for classification purposes, if the frequency separation between zero and non-zero modes is large enough such that they can be spectrally resolved using optical filters before detection, zero-mode lasing can be used for binary classification. For a proof-of-concept-demonstration we consider a $8 \times 8$ array, and we use a freely available database of images of handwriting numbers (details are presented in Sec. 4).

Let us consider an image of a handwritten zero, for instance the one shown in Fig. 1, top left corner, and resize it to a $8 \times 8$ matrix of pixel values, $\boldsymbol{I}^{(1)}$. To encode the image information into the pump profile, $\boldsymbol{P}^{(1)}$, we define an appropriate transformation matrix $\boldsymbol{M}$ such that $\boldsymbol{P}^{(1)} = |\boldsymbol{M}\boldsymbol{I}^{(1)}|$ where the absolute value ensures that the elements of the pump profile are not negative. Then, $\boldsymbol{P}^{(1)}$ is projected onto the nanolaser array by using a SLM. The transformation $\boldsymbol{M}$ has to be chosen such that the resulting the pump pattern $\boldsymbol{P}^{(1)}$ efficiently excites a zero-mode ($\Re[\epsilon_i^{(1)}] = 0$), in such a way that it can reach the lasing threshold, $\Im[\epsilon_i^{(1)}] = 0$ (Fig. 1, right column, top). Note that there can be many zero-modes in the spectrum, because a multiplicity of NH zero-modes can be generated, each one having a different imaginary part. However, in the general case, only one will eventually reach the laser threshold as the pumping is increased [26].

For image classification, we need not only that one particular input image containing a handwritten zero, $\boldsymbol{I}^{(1)}$, leads to lasing zero-mode (i.e., to a *yes* answer), but that any image containing a handwritten zero, $\boldsymbol{I}^{(k)}$, leads to lasing zero-mode (Fig. 1, middle row). Conversely, when the input is not a zero digit, we want that a non-zero mode turns on (Fig. 1, bottom row). Therefore, the key idea is to optimize the coefficients of the transformation matrix $\boldsymbol{M}$ such that a zero-mode turns on if and only if the input image corresponds to a handwritten zero digit.

The following sections describe the machine learning procedure employed, which allows us to obtain the linear transformation, $\boldsymbol{M}$, that optimizes the performance of the binary image classifier.

## 4. Machine learning optimization of the linear transformation matrix

Clearly, the choice of the transformation matrix, $\boldsymbol{M}$, is crucial for obtaining a good classification performance. As previously stated, we are interested in using the zero-modes as detectors of a class of input images, and thus, we want a zero-mode to turn on *if and only if* the pump pattern encodes an image that represents the digit of choice.

We have employed the digit dataset freely available at UCI's ML repository [28], which is a standard dataset for assessing the performance of image recognition systems. We used images that have a resolution that fits the size of the nanolaser array ($8 \times 8$). In order to keep the calculation time reasonably low, we analyzed a subset of 360 images, from which 270 images are used as the *training* set, and 90 images as the *testing* set (see Sec. 4). We aim at training the nanolaser array to recognize a particular digit. Specifically, we can either distinguish between 0s and 1s (*one-vs-one* classifier) or between 0s and any other digit (*one-vs-all* classifier). Both classifiers can be used as building blocks of a multiclass classifier (see Sec. 7).

Typically, the simplest implementation of a machine learning (ML) algorithm for a binary classifier would iteratively adjust its parameters to obtain a good separation between the two populations of the data it is trained with (*training* set). Then, its performance is assessed by looking at new, previously unseen data (*testing* set).

Here we follow this same paradigm. We randomly split the data into two parts: the training set, containing 75% of the samples, and the testing set, with the remaining 25%. The two sets are sampled so that each contains, on average, the same proportion of the two classes of images. In the case of the one-vs-all classifier, we downsampled the number of images containing digits 1 to 9 so the two classes (zero and non-zero digits) are balanced (i.e., zero digits are 50% of the total).

To train the classifier, we have to build a link between its parameters and the classification performance. The idea is to define a smooth cost function correlated with the system errors. In this way, minimizing the cost function improves the performance. The smoothness of the cost function is crucial for designing a numerically stable procedure.

As explained previously, we want a zero-mode to turn on when the input image corresponds to the right class (e.g., a handwritten zero). In addition, we want the gap between the lasing zero-mode and the other modes to be as large as possible, as this will increase the classifier's robustness to noise. Therefore, we define the "spectral gap" of a given image labeled $k$ that is encoded in a pump pattern $\boldsymbol{P}^{(k)}$ and an associated Hamiltonian, $\boldsymbol{H}^{(k)}$, with eigenvalues $\epsilon_i^{(k)}$, as

$$\Delta\epsilon^{(k)} = \max_{i:\left|\Re\left[\epsilon_i^{(k)}\right]\right|\leq\delta} \Im\left[\epsilon_i^{(k)}\right] - \max_{i:\left|\Re\left[\epsilon_i^{(k)}\right]\right|>\delta} \Im\left[\epsilon_i^{(k)}\right], \tag{6}$$

where $\delta$ is a parameter that represents the spectral resolution of the experimental detection system, i.e. the bandwidth of the optical filter that will select a given mode. Ideally $\delta$ is a very small number as the detector should allow to resolve modes with small detuning (see Sec. 5). We will call the *selected modes*, the subset of modes for which $\left|\Re\left[\epsilon_i^{(k)}\right]\right| \leq \delta$. These modes are either zero-modes, or non-zero modes with very small real part.

We note that since at least one mode will lase, either the first or the second term in the r.h.s. of Eq. (6) is null. In particular, if a selected mode lases the first term is 0 and $\Delta\epsilon^{(k)}$ is positive, while it is negative otherwise. The value of the spectral gap depends on the input image and on the linear transformation matrix, $\boldsymbol{M}$. Our goal is to correlate $\Delta\epsilon^{(k)}$ with the image class. For this reason we use the following cost function

$$C = -\sum_{k\in\{+\}} \tanh\left(\eta\Delta\epsilon^{(k)}\right) + \sum_{k\in\{-\}} \tanh\left(\eta\Delta\epsilon^{(k)}\right) \tag{7}$$

where we denote with $\{+\}$ and $\{-\}$ the two sets of images for which we expect the $\Delta\epsilon^{(k)}$ to be positive or negative. Here the hyperbolic tangent represents a differentiable version of the sign function, and $\eta$ is a scale parameter which has been fixed to $\eta = 4$ after few experiments (the system's performance is rather unaffected by $\eta$).

## 5. Algorithm implementation

The cost function depends on the elements of $\boldsymbol{M}$ through the values of $\Delta\epsilon^{(k)}$, and it is likely to display several local minima. For this reason, we employ a dual annealing minimization of $C$ with respect to $\boldsymbol{M}$ [29]. Once the cost function was minimized using the images of the training set, the matrix $\boldsymbol{M}$ obtained was used to operate the classifier and compute its performance on both, training and testing sets.

The pipeline to classify an input image is composed of the following steps:
1) The image $\boldsymbol{I}^{(k)}$ is encoded in a pump pattern as $\boldsymbol{P}^{(k)} = |\boldsymbol{M}\boldsymbol{I}^{(k)}|$.
2) The pump values are adjusted as $\boldsymbol{P}^{(k)} \to \alpha\boldsymbol{P}^{(k)}$, where $\alpha$ is a constant that is selected such that only one mode, say mode $j$, reaches laser threshold, i.e. $\Im[\epsilon_j^{(k)}(\alpha)] = 0$. This requires to find $\alpha$ for which $\max_i \left\{\Im[\epsilon_i^{(k)}(\alpha)]\right\} = 0$, that is a root-finding problem. To improve performance,

we replace the max with a differentiable *softmax* function $\sigma^{(k)}(\alpha)$:

$$\sigma^{(k)}(\alpha) = \log \left( \sum_{i=1}^{64} \exp \left( \Im[\epsilon_i^{(k)}(\alpha)] \right) \right). \quad (8)$$

3) Calculate the spectrum of the Hamiltonian, pumped by the adjusted pump.

4) Calculate the "spectral gap" (Eq. 6) and obtain the result of the classification task from the sign of the gap: if $\Delta\epsilon^{(k)} > 0$ the lasing mode is a selected mode (i.e., it is either a zero-mode, or a non-zero mode with very small real part) and the answer is yes, else, the lasing mode is a non-selected mode and the answer is no.

From the numerical point of view, the most demanding part of the pipeline is to calculate $\alpha$ to adjust the pump. The reason is that every evaluation of the softmax function at a different $\alpha$ implies the resolution of an eigenvalue problem. The cheapest operation of the pipeline is the calculation of the pump previous to the adjustment. We stress that this is the only operation that in an experimental setup would have to be done numerically. All the other steps would come "for free" from the dynamics of the nanolaser array, performed optically at the hardware level. In addition, as will bediscussed in Sec. 7, the matrix multiplication could also be implemented all optically using a passive 2D metasurface.

The optimization was run until a maximum number of evaluations of the cost function ($\sim 3 \times 10^5$) was reached. On a 40 cores cluster, the optimization typically takes about 24 hours.

The calculations were done with the following re-scaled parameters: $\gamma = 0.2$ and $\kappa_x = \kappa_y = 1$. The former leads to a cavity dissipation rate of $\tau^{-1} = \pi c \Delta \lambda / \lambda^2 = 157$ GHz (cavity damping time $\tau = 6.4$ ps), while the latter leads to a coupling rate of $5\tau^{-1}$. Both parameters are consistent with standard evanescently coupled L3 photonic crystal nanlasers separated by three rows of holes in the $\Gamma M$ direction, featuring a mode splitting of 5 times the cavity linewidth; such a splitting is a typical value, which can be modified if needed by means of the barrier engineering technique [22, 23]. Importantly, we have chosen the coupling strength large enough in order to overcome the intercavity detunings that may arise as a consequence of spatial inhomogeneities due to fabrication imperfections (RMS of about a resonance linewidth). Concerning the spectral cutoff ($\delta$-parameter), it is related to the bandwidth of the optical passband filter that will select laser modes at the output layer. Passband widths in commercial tunable bandpass filters in the telecommunications band can be as small as $\Delta \nu = 10$ GHz which, in normalized units, leads to $\delta = 2\pi T \Delta \nu \approx 0.08$. Hence, only $\delta$ values above 0.08 can be considered as realistic bandwidths in experiments. We will then analyse two cases: $\delta = 0.1$, compatible with realistic applications, and $\delta = 10^{-3}$ (ultra-narrow filter) for a deeper understanding of the protocol and comparison.

## 6. Results

Table 1 summarizes the metric scores obtained by the two image classifiers, one-vs-one and one-vs-all. As usual, we define the *accuracy* as the fraction of correct classifications over the total number of images in the train/test dataset, the *recall* as the fraction of 0s in the dataset that were correctly identified, and the *precision*, as the fraction of correctly predicted 0s over the number of predicted 0s.

### 6.1. One-vs-one classifier

The one-vs-one classifier is, on average, the most precise. In the case of the ultra-narrow filter, if we simply look at the performance metrics, we have the best classifier; the algorithm only fails on one example of the test set.

In Fig. 2 we report two examples of classification for a 0 and of a 1. As we can see, the images are correctly classified, but the difference in frequency between the zero-mode and the non-zero-mode is very small. To understand this small difference, we plot in Fig. 3(a) the distribution of the frequencies of the lasing modes, for the whole image dataset. We can identify two main regions: one at extremely low real parts —limited to the floating point numerical resolution— that contains the zero-modes, and another one between $10^{-4}$ and $10^{-1}$. The latter has a bimodal distribution, with a main peak above the selection threshold and a minor secondary peak right below. So the first observation is that some selected modes have small frequencies but they are not strictly zero-modes. To quantify their fraction with respect to the zero modes, we report in Fig. 3(b) the cumulative distribution functions (CDFs) of the lasing frequencies for the selected and for the non-selected modes. We can see that, when the input image is a zero digit, ∼5% of the selected modes are not strictly zero-modes but they are very close to the detection cutoff. Moreover, also a small fraction of the non-selected modes close to the selection boundary also come from zero input digits: only one for $\delta = 10^{-3}$ (Fig. 3a), and 5 for $\delta = 0.1$ (Fig. 3c), slightly lowering the recall value (see Table 1). As a result, the training precision is 100% for the one-vs-one classifier regardless the $\delta$-value, meaning that all identified zeroes originally come from handwritten zero inputs.

|  |  | One-vs-one | | One-vs-all |
|---|---|---|---|---|
| Resolution | $\delta$ | $10^{-3}$ | $10^{-1}$ | $10^{-1}$ |
| Accuracy | *Train* | 100% | 98.9% | 97.8% |
|  | *Test* | 98.9% | 97.8% | 96.7% |
| Precision | *Train* | 100% | 100% | 97.7% |
|  | *Test* | 100% | 100% | 97.9% |
| Recall | *Train* | 100% | 97.6% | 97.7% |
|  | *Test* | 98% | 96.1% | 95.8% |

Table 1. Summary of the scores obtained for one-vs-one and for one-vs-all classifiers.

Clearly, classification in our optical machine reduces to a linear boundary separation problem given by a frequency cutoff $\delta$. The smaller the cutoff, the better the separation between classes. Specifically, 100% accuracy is obtained with $\delta = 10^{-3}$ for the training set, while it slightly drops for the test set (Table 1). However, increasing the cutoff to $\delta = 0.1$, the accuracy keeps high but it is slightly smaller (98.9%) in the training test, meaning that a larger fraction of "zero" input images are incorrectly classified as "ones".

It is important to point out that, although the metrics are slightly worse in the case of the wide filter ($\delta = 0.1$), they prove to be more robust. In Fig. 4 we can see two examples of correct classifications. The non-selected lasing mode is significantly far from the filter cutoff (Fig. 3), while ∼4% of the selected modes within the range $\Re(\epsilon) \sim [10^{-3}, 10^{-2}]$ are not strictly zero modes, still reasonably far from the $\delta = 0.1$ cutoff.

### 6.2. One-vs-all classifier

The one-vs-all classifier is, on average, less accurate than the one-vs-one classifier, a fact that should not be surprising because the variance of the "negative" examples is higher. Similarly to the wide-filter one-vs-one classifier, there is a good separation between the two classes: zero digits exciting zero and near-zero modes, and non-zero digits leading to non-zero lasing modes (see Fig. 5). Unlike the one-vs-one classifier, here the precision is below 100%, with a small though nonzero fraction of other digits that trigger —both zero and non-zero— selected modes, and consequently they are incorrectly classified as "zeros", as it can be seen in Fig. 3(e). On the other hand, the higher diversity of the negative examples affects the fraction of selected modes

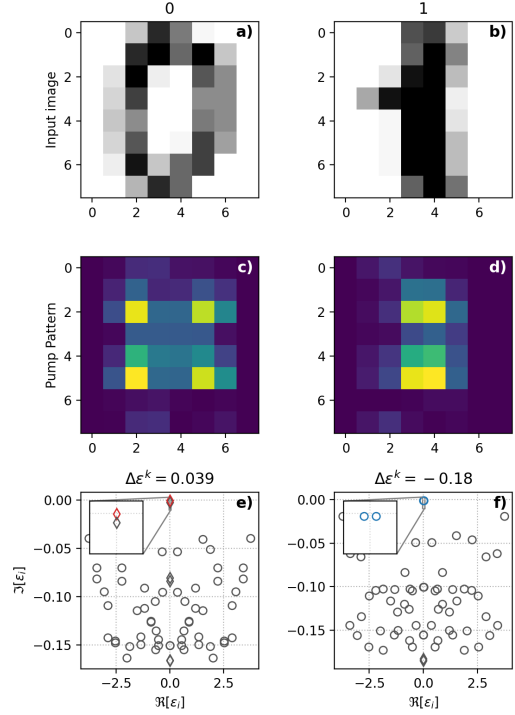

Fig. 2. Two examples of correct image classification for the one-vs-one classifier. **a, b)** Input images. **c, d)** The corresponding pump patterns, computed as $P = |M\,I|$ using the $M$ matrix that minimizes the cost function defined in Eq. (7). **e, f)** Real and imaginary part of the spectrum of the Hamiltonian matrix pumped with the patterns depicted in c) and d) respectively. In red the lasing eigenvalues that are selected modes, that is the eigenvalues for which $-\delta \leq \Re[\epsilon] \leq \delta$ with $\delta = 10^{-3}$. In blue the lasing eigenvalues that are not selected modes. Diamonds represent zero-modes, while circles represent non-zero modes. The two insets range from -0.05 to 0.05 on the horizontal axis and from -0.01 to 0.003 on the vertical axis.

that are not zero-modes which, in this case, is ∼20%, as it can be seen in Fig. 3(f).

### 6.3. Robustness to noise

To assess the robustness of our classifier to noise, we perturbed the input images with a spatially uncorrelated uniformly distributed random intensity. We normalize the noise level by the maximum pixel value in the original image, meaning that a noise level of 1 is a nearly random image. We then used the trained algorithm to classify the perturbed images and computed the accuracy degradation as the noise level grows.

The results are depicted in Fig. 6. Note that the one-vs-one classifier is more robust when setting $\delta = 10^{-1}$ compared to $\delta = 10^{-3}$. This is consistent with our previous observation the frequency gap between selected and non-selected modes is wider in the former case. This confirms our previous statement that larger cutoff detection filters are more robust despite being slightly less precise.

The one-vs-all classifier is less robust to noise compared with the one-vs-one classifier having the same $\delta$-value. In all the cases, there is an abrupt drop in accuracy as the noise intensity is increased, and the accuracy asymptotically tends to random (50%) for large noise intensities. For the one-vs-one classifier with a wide filter, the accuracy is tolerant to about 20% of relative noise intensity. Overall, we can state that if the noise intensity is below 5% of the maximum image intensity, the classifier's performance is almost unaffected.

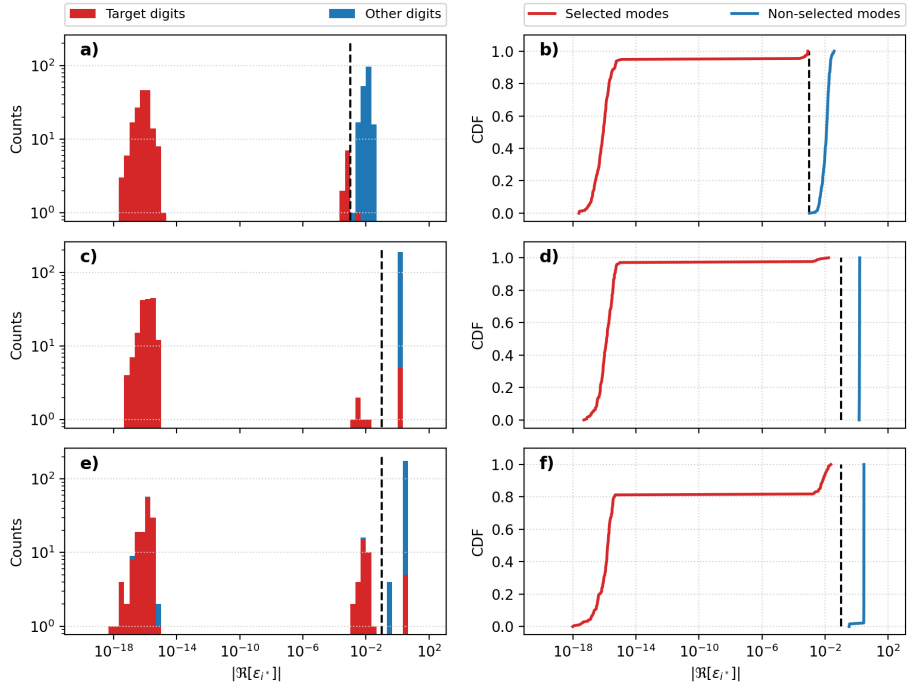

Fig. 3. **a)** Histograms of the real parts of the lasing modes $\epsilon_{i^*}$, where $i^*$ is the particular eigenvalue at lasing threshold, for all the images of the dataset. The bin size increase with $\Re[\epsilon_{i^*}]$, in order to have equal bin size in log scale. Counts are also reported in log scale. In red, eigenvalues corresponding to the target digit, "0", while in blue we report the eigenvalues corresponding to any other digit. **b)** Cumulative distribution functions (CDFs) of $\Re[\epsilon_{i^*}]$ for all the images, for the one-vs-one classifier. In red the CDF of the selected modes, that is the lasing eigenvalues for which $-\delta \leq \Re[\epsilon_{i^*}] \leq \delta$ with $\delta = 10^{-3}$. In blue the distribution of the remaining lasing modes. Both in a) and b) the value of the cutoff $\delta$ is depicted as a black vertical dashed line. **c), d)** Same as a),b) but for $\delta = 0.1$. **e), f)** Same as c),d) but for the one-vs-all classifier.

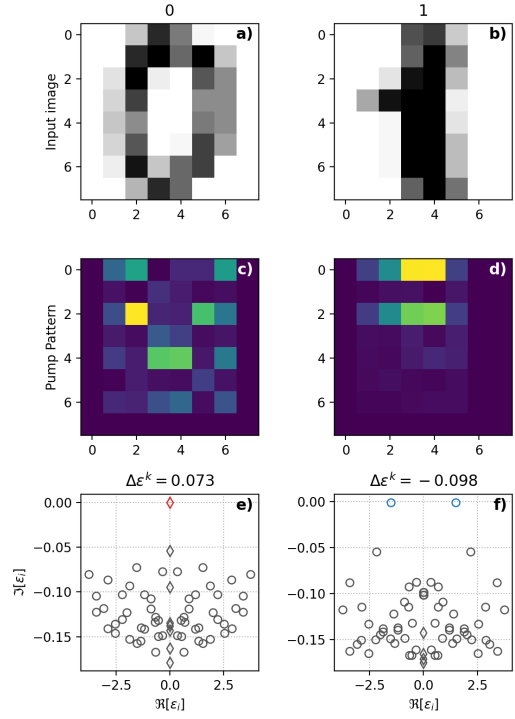

Fig. 4. As Fig. 2 but for $\delta = 10^{-1}$.

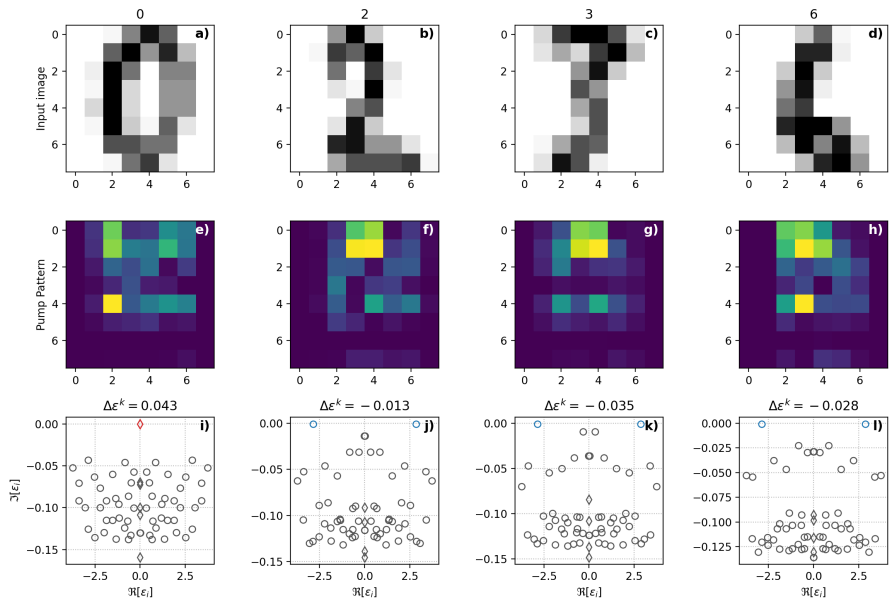

Fig. 5. As Fig. 4 but for the one-vs-all classifier.

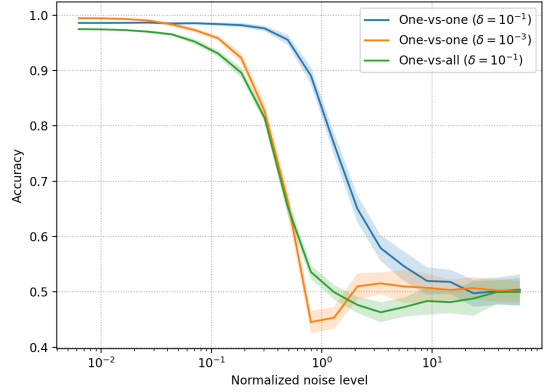

Fig. 6. Accuracy of the classifiers as a function of the noise level. The noise intensity is normalized by the maximum pixel value in the original. The shaded area represent the standard deviation of the accuracy values over 20 independent realizations.

## 7. Discussion

We have demonstrated that the modes emitted by a nanolaser array can be used for binary image classification, being able of both, one-vs-one and one-vs-all classification. The selected modes detecting the positive class are mainly zero-modes. A fraction of the selected modes, however, does not correspond to zero-modes and this fraction is significant in the case of the one-vs-all classification. We checked whether their eigenstates are similar to those corresponding to zero-modes ($\phi/2$-phase difference between neighbour cavities [26, 27]), and we found that there is a certain similarity only if the non-zero modes are coalescing into a zero-mode. In the case of the one-vs-one classification, these small clusters have small real parts because they are either coalescing into zero-modes or they have a zero real part just below the lasing threshold. For this reason, the pump pattern is remarkably similar to zero-mode pumps and they can correctly classify zero digits. Using a very narrow filter would force the classification to rely on zero-modes solely. Thus, we tested even smaller values of $\delta$ ($\delta < 10^{-3}$), but obtained a decrease in the accuracy. Therefore, in order to obtain a good performance, small clusters of non-zero modes must be included in the classification. It might be possible to eliminate these small clusters (and use only the zero-modes for yes/no classification) by changing the coupling strength between the cavities. Therefore, it will be interesting in the future to study the relationship between the system's parameters and the spectral mode distribution.

We have proposed an ANN for image classification that operates mainly at the hardware, optical level, with the only in-silico operation being a matrix multiplication, which is extremely fast and highly parallelizable. We are now searching alternatives to this step, such as 2D passive metasurfaces, to realize all-optically the in-silico MAC functions. These metasurfaces, similar to the ones reported in Ref. [5] but in the optical domain, would take the form of external phase masks that can be fabricated with standard silicon or dielectric nanotechnology.

Beyond binary classification, a multi-class classifier can be built by using binary classifiers and an aggregation rule [30]. With one-vs-one classifiers, having $N$ classes, we will need $N(N-1)/2$ classifiers, while with one-vs-all classifiers, we only need $N$ classifiers. The predictions are aggregated and a final prediction is made, and confidence scores can be assigned to each class. Moreover, it is possible to design a 2D nanocavity array that supports a set of topologically protected quasi-zero modes. Specifically, we have conceived a 2D extension of a SSH model [31, 32], where the quasi-zero-modes formed by topological defects can exhibit different frequencies. Consequently, all the defect modes are highly localized in terms of their spatial distribution and lasing frequencies, and as a result these topologically protected modes may prove to be mutually independent. Such a system, which goes beyond the scope of the present work and will be investigated elsewhere, might enable N-class optical classifiers based on topologically protected laser modes.

We stress that our aim is not to overcome the performance of state-of-the-art in-silico classifiers, as a perceptron or a random forest can perform very well on this data. Instead, our goal is to show that a nanolaser array can serve to implement an all-optical multilayer ANN device, a longstanding goal in optical computing. As a matter of fact, as explained in Sec. 2, by virtue of the sublattice decomposition, the 2D nanocavity array with nearest-neighbor couplings can actually be mapped to a bi-layer system —the checkerboard—, without any intralayer coupling. Moreover, because of in-situ gain-saturation nonlinearity, the activation function is applied on each layer. Depending on the array geometry and couplings, the number of hidden layers could increase from 2 to 4. Because of the non-local character of the Fourier transform, there is potential for global connectivity between the input layer (the pump pattern that encodes the input image) and the hidden layers (the nanocavities), and also between the hidden layers and the output layer (the modes of the array). Therefore, even though the layer-to-layer connectivity is local, the input and output layer connections are global. This architecture, used here for a proof-of-concept demonstration, has given high accuracy for image classification of handwritten digits.

The resolution of the images that can be processed is limited by the size of the nanolaser array; however, in many applications there is no need of high image resolution. Note that, in an all-optical pipeline, the pre-processing convolution step, in the form of image pixelization, can be implemented by spatial filtering. Finally, we would like to point out that the eigenvalue problem is an idealized representation of a laser array device, for small enough pumping rates and as long as a single mode eventually reaches the threshold. However, in a real laser device, once a given mode —such as the zero one— reaches the threshold, further increasing the pump power will not make another laser mode turn on, since the gain will remain clamped at the loss level of the first above-threshold lasing mode. Otherwise saying, because of gain saturation, single mode operation is likely close enough to the laser threshold in a homogeneously broadened gain medium such as III-V semiconductors, and provides high mode selectivity mechanisms that can be used for machine learning purposes.

## 8. Conclusions

We have proposed an optical computing system based on a bidimensional array of semiconductor nanolasers. An optical pump beam is spatially modulated through a computer-based linear transformation applied to a SLM, in such a way that only a given class of input images efficiently excites a zero-mode of the array. With this protocol, overall accuracies of 98% for one-vs-all classifiers have been obtained. In further device conceptions, the in-silico pre-formatting MAC operations could be replaced by all-optical transformations using 2D metasurfaces.

Because of sublattice symmetries and in-situ gain saturation nonlinearities of the nanolaser array, our system is a promising candidate for an all-optical multilayer neural network device, potentially benefiting from the low energy consumption and fast light-matter interactions in semiconductor nanocavities.

The originality of our device relies on: i) complex free-space operations such as diffraction, convolutions and mode decomposition (matrix diagonalization) that, in optics, are carried out "for free"; ii) a low-energy-consumption device as a nanolaser, that can saturate with few photons (high $\beta$-factor thresholdless laser regime); iii) III-V semiconductor coupled cavity network geometries, which provide both multilayer architectures with optical coupling engineering and in-situ saturation nonlinearities in a natural way; iv) nonlocal activation functions because of mode coupling which, as is has been investigated in Ref. [25], improves the classification (form nonlinear to linear regression); v) the symmetry —or even topological— protection of laser modes. The latter is particularly interesting, since it seems to provide a natural separation boundary for classification, in close analogy with invariant manifolds in a nonlinear dynamical phase space. Therefore, we believe that our nanolaser array classifier will open new and promising avenues of research in optical machine learning.

## Disclosures

The authors declare no conflicts of interest.

## Data Availability Statement

The digit image dataset is freely available at UCI's ML repository [28].

## Funding

This work is supported by a public grant overseen by the French National Research Agency (ANR) as part of the "Investissements d'Avenir" program (Labex NanoSaclay, reference: ANR-10-LABX-0035) and by the ANR UNIQ DS078. G. T. and C. M. are supported in part by Ministerio de Ciencia, Innovación y Universidades (PGC2018-099443-B-I00); C. M. also acknowledges funding from Institució Catalana de Recerca i Estudis Avançats (Academia). K. J. acknowledges the support from China Scholarship Council (No.202006970015).